\documentclass[twocolumn,doublespacing,showpacs,preprintnumbers,amsmath,amssymb]{revtex4}
\usepackage{amssymb}
\usepackage{txfonts}

\usepackage{graphicx}% Include figure files
\usepackage{subfigure}
\usepackage{dcolumn}% Align table columns on decimal point
\usepackage{bm}% bold math
\usepackage{sidecap}

\makeatletter
\def\@begintheorem#1#2{\trivlist
  \item[\hskip \labelsep{\bfseries #1\ }]\itshape}
\makeatother

\begin{document}

\title{Passive scheme analysis for solving untrusted source problem in quantum key distribution}% Force line breaks with \\

\author{Xiang Peng}\thanks{E-mail: xiangpeng@pku.edu.cn.}
\author{Bingjie Xu}
\author{Hong Guo}\thanks{E-mail: hongguo@pku.edu.cn.}
\affiliation{%
CREAM Group, State Key Laboratory of  Advanced  Optical
Communication Systems and Networks (Peking University) and Institute
of Quantum Electronics, School of Electronics Engineering and
Computer Science, Peking University, Beijing 100871, China}

\date{\today}

\begin{abstract}
As a practical method, the passive scheme is useful to monitor the photon statistics of an untrusted source
in a ¡°Plug \& Play¡± quantum key distribution (QKD) system. In a passive scheme, three kinds of monitor mode
can be adopted: average photon number (APN) monitor, photon number analyzer (PNA), and photon number
distribution (PND) monitor. In this paper, the security analysis is rigorously given for the APN monitor, while for
the PNA, the analysis, including statistical fluctuation and random noise, is addressed with a confidence level.
The results show that the PNA can achieve better performance than the APN monitor and can asymptotically
approach the theoretical limit of the PND monitor. Also, the passive scheme with the PNA works efficiently when
the signal-to-noise ratio ($\rm {R^{SN}}$) is not too low and so is highly applicable to solve the untrusted source problem in
the QKD system.
\end{abstract}

\maketitle

\section{INTRODUCTION}
\begin{figure}[t]
\centerline{\includegraphics[width=3.5in]{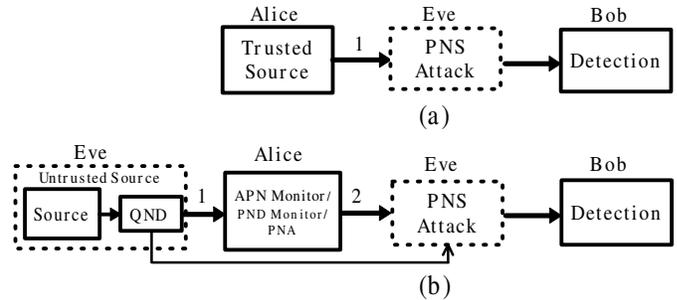}} \caption{(a)
Scheme of ``trusted source''. (b) Scheme of ``untrusted source''.}
\label{Fig:Trust_Untrust}
\end{figure}

Quantum key distribution (QKD) establishes two parties (Alice and
Bob) to share a secure key
\cite{BB_84,Ekert_91,RevQKD_GRTZ_02,RevQKD_Dusek_06,RevQKD_Zhao_08,RevQKD_Lut_08}.
The single-photon BB84 protocol \cite{BB_84} for ideal
\cite{LoChauQKD_99,ShorPreskill_00,Mayers_01} and practical
(imperfect single-photon source, channel and detection)
\cite{ILM_07,GLLP_04} QKD systems has proved to be unconditionally
secure in the last decade.  To efficiently apply the security
analysis of \cite{ILM_07,GLLP_04}, it is better that the
photon-number distribution (PND) of the source is fixed and known to
Alice and Bob, while Eve cannot control and change it. This kind of
source is defined as a ``trusted source" [see
Fig.~\ref{Fig:Trust_Untrust}(a)]. Due to channel loss and the
multiphoton states of the trusted source, Eve can perform the
photon-number-splitting (PNS) attack
\cite{IndividualAttack_00,BLMS_00,LutkenhausJahma_02} without
causing any disturbances and obtain full information from the  keys
generated by the multiphoton states. Thus, all the losses and
errors are pessimistically assumed from the single-photon state of
trusted source and the secure key rate of QKD is reduced.
Fortunately, with the decoy-state method
\cite{Hwang_03,Wang_05,Decoy_05,Wang2_05,Practical_05}, the
properties of the quantum channel are characterized by Alice and
Bob, thereby higher secure key rate can be achieved
\cite{RevQKD_Lut_08}, which has been successfully implemented in
experiments for QKD with trusted source
\cite{decoy_Rosenberg_07,Zeilinger_Decoy_07,PanDecoy_07,YSS_Decoy_07}.

However, it was recently  found that the characteristics of the QKD source need to be verified in a real-life experiment
\cite{Gisin_06,Wang_07_source,Wang_Fluc_PRA_07,Wang_08_source,Wang_09_source,ZQL_untru_08,Peng_untrusted_08,Moroder_untrusted_08,Curty_Source_09,Adachi_Source_09,Xu_Fluctuation_09}.
For example, the intensity fluctuation from the source makes the
assumption of the trusted source fail in decoy-state protocol, for which
a rigorous security analysis has been given
\cite{Wang_07_source,Wang_08_source,Wang_09_source}. Especially, the
assumption of a trusted source does not hold in the round-way ``Plug \&
Play'' system \cite{PaP_Gisin_02}, in which Bob sends to Alice a
train of bright laser pulses which can be eavesdropped and
controlled by Eve. Even though Alice can use a time (frequency)
-domain filter and phase randomizer \cite{Zhao_07} to assure the
mixture of single-mode Fork states
\cite{Gisin_06,ZQL_untru_08,Peng_untrusted_08}, the PND of
classically mixed states is under Eve's control provided that Eve
has the power of  a quantum nondemolition (QND) measurement
\cite{WallsMilburn_94} and then she can know the photon number of
each pulse sent to Alice's station. This kind of  source is defined
as an ``untrusted source" [see Fig.~\ref{Fig:Trust_Untrust}(b)]. Note
that Alice's filters, phase randomizer and encoder exist in Alice's
side which are not shown in Fig.~\ref{Fig:Trust_Untrust}(b). Through
QND in Fig.~\ref{Fig:Trust_Untrust}(b), Eve knows the exact photon
number at position 1, which may assist her PNS attack at position 2
\cite{ZQL_untru_08}.

For applying the BB84 protocol in the experiment, an average photon number
(APN) monitor for an untrusted source was proposed \cite{PaP_Gisin_02}.
However, until now no quantitative and/or detailed analysis to
prove the effectiveness of the method was reported. For keeping the
efficient decoy-state analysis for an untrusted source, it was proposed
to estimate the lower and upper bounds of a few parameters about the PND
at position 2 in Fig.~\ref{Fig:Trust_Untrust}(b)
\cite{Wang_07_source,Wang_08_source,Wang_09_source}. Recently, the
detector-decoy scheme was theoretically proposed to monitor the PND
of untrusted source using a threshold detector
\cite{Moroder_untrusted_08}. From another viewpoint, an active
scheme of the photon number analyzer (PNA) is put forward
\cite{ZQL_untru_08} even though it is hard to put into reality
\cite{Peng_untrusted_08}. In a recent work, a passive scheme of PNA
was proposed and experimentally tested though
some practical issues (e.g., statistical fluctuation and detection
noise) were not considered \cite{Peng_untrusted_08}.

In the following, the security analysis is made for the APN monitor and
the PNA. For the PNA, some practical issues, such as statistical fluctuation
due to a finite number of measurements (estimated by the Clopper-Pearson
confidence interval \cite{Clopper_Pearson_CI_34,Larsen_CI_86}) and
two kinds of additive detection noise (Poissonian and Gaussian
electronic noise) are analyzed. It shows that  the PNA has better
enhancement in a secure key rate than the APN monitor and so is more
applicable in BB84 protocol.

\section{security analysis for the APN monitor}\label{APN_security_analysis}
\begin{figure}[t]
\centerline{\includegraphics[width=2.5in]{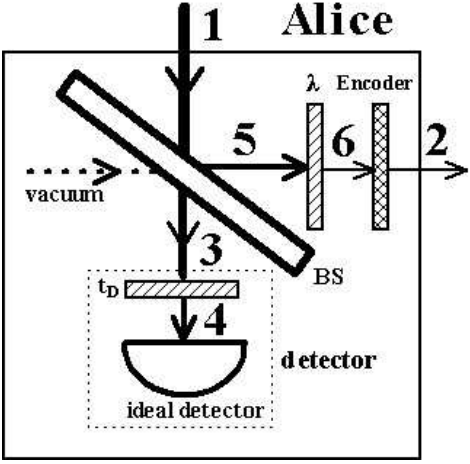}} \caption{The
passive scheme using a beam splitter (BS, transmittance: $t_{B}$), a
detector (efficiency: $t_{D}$), an attenuator (transmittance:
$\lambda$), and an encoder. Here, the imperfect detector is modeled
by a virtual attenuator between P3 and P4 (transmittance: $t_D$) and
an ideal detector (efficiency: $100\%$). The encoder is used for
phase coding.} \label{Fig:BS_Det}
\end{figure}

A passive scheme is illustrated  in Fig.~\ref{Fig:BS_Det} where the
detector  is limited to monitor the APN ($\mu$) of the untrusted
source and an attenuator (transmittance: $\lambda$) is used to
ensure the weak pulses with the fixed APN (such as 0.1). This simple
intensity monitor has been implemented to a practical ``Plug \&
Play'' system \cite{PaP_Gisin_02}. Applying  the BB84 protocol without
the decoy states, the security analysis of the untrusted source is
presented below. For convenience, in the following discussion, P$i$
with $i =1, \ldots$, refers to the position $i$.
In Fig.~\ref{Fig:BS_Det}, the photoelectron number distribution
$D(m)$ at P4 and the PND $P(n_2)$ at P2 are the Bernoulli transform
of PND $P(n_1)$ at P1, that is
\begin{eqnarray}
D(m)&=& B[P(n_1),\xi]=\sum_{n_1=m}^{\infty}P(n_1){n_1 \choose
m}\xi^{m}(1-\xi)^{n_1-m}\label{Untru:PnDm},\\
P(n_2)&=&B[P(n_1),\eta]=\sum_{n_1=n_2}^{\infty}P(n_1){n_1 \choose
n_2}\eta^{n_2}(1-\eta)^{n_1-n_2}\label{Untru:PnPn},
\end{eqnarray}
where $\eta=\lambda(1-t_B)$ and $\xi=t_Bt_D$. In practice, the
intensity monitor of the detector only gives $\langle
m\rangle=\sum_{m=0}^\infty mD(m)$. However, based on
Eq.~(\ref{Untru:PnDm}), $\mu=\langle n_1\rangle=\langle
m\rangle/\xi$ is known if $\langle m\rangle$ and $\xi$ are both
exactly known. Then, from Eq.~(\ref{Untru:PnPn}), the APN $\langle
n_2\rangle$ ($=\eta\mu$) at P2 can  be derived. From the work of
GLLP, BLMS: see authors of  Refs. \cite{GLLP_04,BLMS_00}, Alice should estimate
the upper (lower) bound of multiphoton (single-photon) probability
at P2. For the untrusted source only under Alice's APN monitor, Eve
can arbitrarily manipulate $P(n_1)$  with the constrains of
$\sum_{n_1=0}^\infty n_1P(n_1)=\mu$ and $\sum_{n_1=0}^\infty
P(n_1)=1$. Eve chooses the optimal $P(n_1)$ to maximize  the
multiphoton probability at P2 for eavesdropping more information by
the PNS attack. Alice has to estimate the worst case and the upper bound
of the multiphoton probability is estimated at P2, which is
\begin{equation}\label{Untru:MultiPhoton}
\begin{aligned}
P(n_2>1)&=\sum_{n_1=2}^{\infty}P(n_1){n_1 \choose
2}\eta^{2}(1-\eta)^{n_1-2}+\cdots\\
&=a_2P(n_1=2)+\cdots+a_kP(n_1=k)+\cdots,
\end{aligned}
\end{equation}
where $a_k=1-(1-\eta)^k-k\eta(1-\eta)^{k-1}~(k\geq2)$. Thus, to
estimate the worst case, one needs to solve the convex optimization
\cite{Boyd_04} or \emph{linear program} (LP) problem which has the
form
\begin{equation}\label{Untru:convex_optimization}
\begin{aligned}
\rm{minimize} ~&-c^Tx,\\
\rm{subject~to}~&Ax=b,~x\geq0,
\end{aligned}
\end{equation}
where
\begin{eqnarray}\label{Untru:convex_optimization_para}
&&x=\left[\begin{array}{llllll}P(n_1=0)&P(n_1=1)&P(n_1=2)&\cdots&P(n_1=k)&\cdots\end{array}\right]^T,\nonumber\\
&&c^T=\left[\begin{array}{llllll}0&0&a_2&\cdots&a_k,&\cdots\end{array}\right],\nonumber\\
&&A=\left[ \begin{array}{llllcl}
0 & 1 & 2 & \cdots & k & \cdots\nonumber\\
1 & 1 & 1 & \cdots & 1 & \cdots\nonumber\\
\end{array} \right],\nonumber\\
&&b=\left[ \begin{array}{l}
\mu \\
1\\
\end{array} \right].
\end{eqnarray}
The LP problem  can be solved by using the \emph{simplex method}
\cite{Press_NumericalRecipes_86} and the maximum value
$\overline{P}_{n_2>1}$ of $c^Tx$ is given as
\begin{equation}\label{APN_monitor_upperbound}
\overline{P}_{n_2>1}=\frac{a_{k_s}\mu}{k_s},
\end{equation}
where $a_{k_s}/k_s$ is the maximum value of $a_k/k$, and the optimal
$P(n_1)$ of the untrusted source for Eve has the form
$\left[P(n_1=0)=1-\mu/k_s,P(n_1=k_s)=\mu/k_s,P(n_1\neq0,k_s)=0\right]$.
For example, if $\eta=0.001$ and $\mu=100$ are fixed, the \emph{simplex
method} gives that $\overline{P}_{n_2>1}=0.02985$ when
$P(n_1=0)=1-\mu/1794$, $P(n_1=1794)=\mu/1794$, and
$P(n_1\neq0,1794)=0$.

For an error-free setup with Bob's perfect detection ($100\%$
detection efficiency and no dark counts), the necessary condition
for security is \cite{BLMS_00}
\begin{equation}\label{security_necessary}
\overline{\Delta}=\frac{\overline{P}_{n_2>1}}{Q_{e}}<1,
\end{equation}
where $Q_{e}$ is the total expected probability of detection events and
$\overline{\Delta}$ is the upper bound of tagged signal probability
\cite{GLLP_04}. Otherwise, Eve can suppress the single-photon
signals completely and obtain full information on the multiphoton
signals. In the ``Plug \& Play'' system, the expected photon source
entering Alice's side is Poissonian with APN $\mu$ (i.e.,
$Q_{e}=1-e^{-\mu\eta\eta_f}$). Here, $\eta_f=10^{-\alpha' L/10}$ is
the transmittance of communication fiber with the loss $\alpha'$
(0.21 dB/km@1550 nm) and the length $L$. With
$\mu=100,~\eta=0.001,~\alpha'=0.21$, and
$\overline{P}_{n_2>1}=0.02985$, by Eq.~(\ref{security_necessary}),
the secure transmission distance $L<24.7$ km. Note that, in the same
setup, if Alice successfully monitors the PND of the untrusted source at
P1 or P2 and Eve does not replace the Poissonian source, the secure
transmission distance $L<63$ km. Generally, for a practical ``Plug
\& Play'' setup with quantum bit error rate (QBER) $E_{e}$, the
secure key rate for an untrusted source with the APN monitor is
\begin{equation}\label{APN_GLLP}
R\geq\frac{1}{2}Q_{e}\left\{-f\left(E_{e}\right)H_2(E_{e})+(1-\overline{\Delta})\left[1-H_2\left(\frac{E_{e}}{1-\overline{\Delta}}\right)\right]\right\},
\end{equation}
where $f(E_{e})H_{2}(E_{e})$ is the leakage information in the error
correction and $H_{2}(x)=-x\log_{2}x-(1-x)\log_{2}(1-x)$. From the
above analysis, one cannot find the secure key rate through 67 km fiber
in \cite{PaP_Gisin_02}.

The APN monitor does not require high detector resolution.  In
the experiment, an optical power meter records the time average of the total
pulse energy during one fixed period and is commonly used to monitor
the mean optical power or APN $\langle m\rangle$. Due to finite
average time, the measured values of the power meter may statistically
fluctuate between different periods. Thus, one of the records from
the power meter cannot represent the real $\langle m\rangle$ unless
it has an infinite average time. However, when the running time of the QKD
system is much longer than the average time of the power meter and the large
number of records from the power meter are obtained, approximately,
the mean of these records obeys the normal distribution according to
the central limit theorem (CLT) \cite{Papoulis_01}. Through the mean and
variance of these records, the real $\langle m\rangle$ can be
statistically estimated in an interval $[\langle m\rangle^L,\langle
m\rangle^U]$ with a confidence level \cite{Papoulis_01}. Further,
with the same confidence level, $\mu\in[\mu^L,\mu^U]$ can be
estimated, where $\mu^L=\langle m\rangle^L/\xi$ and $\mu^U=\langle
m\rangle^U/\xi$. From Eq.~(\ref{APN_monitor_upperbound}),
$\overline{P}_{n_2>1}$ can be estimated by $a_{k_s}\mu^U/k_s$.

\section{security analysis for the PNA} \label{PNA_security_analysis}
The passive setup in Fig.~\ref{Fig:BS_Det} can realize the PNA which
leads to different analysis results \cite{ZQL_untru_08}. By
replacing the parameter $\lambda$  and randomly choosing the vacuum
state and the signal (weak decoy) state with $\lambda_s$
($\lambda_d$), the three-intensity decoy-state protocol can be applied
\cite{Peng_untrusted_08}.

The PNA needs to estimate the lower bound of the fraction $1-\delta$ of
the photon pulses (defined as ``untagged bits'' and are originally
referred to as P1 in Fig. 2 \cite{ZQL_untru_08}), and the photon number
$N$ of which falls in the preset range of $[N_{\rm{min}},N_{\rm{max}}]$.  To
estimate $1-\delta$, a detector needs to monitor the PND at P4,
which is used to yield that at P1 using the inverse-Bernoulli transform
\cite{Peng_untrusted_08}. Another important parameter to estimate
the secure key rate is the transmittance $\lambda^A$ for ``untagged
bits'' in Alice's side [for those at P1, $\lambda^A=(1 -
t_B)\lambda$].

\begin{figure}[t]
\centerline{\includegraphics[width=3in]{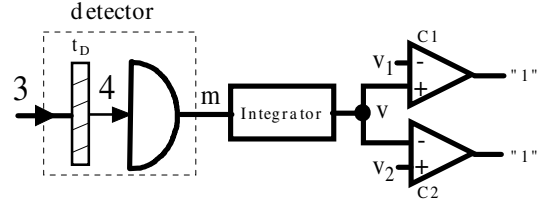}} \caption{A
two-threshold detection for the PNA. The integrator converts the
photoelectron number $m$ to the voltage $v$. C1 (C2): comparator.
``11" output from C1 and C2 means the voltage $v$ between $v_1$ and
$v_2$ and the photon number $m$ falls in the range $[m_1,m_2]$,
correspondingly.} \label{Fig:PNA_detector}
\end{figure}

The effectiveness of inverse-Bernoulli-transform algorithm is
sensitive to the statistical fluctuation and the detection noise,
and high-resolution detection is required. To implement the passive
scheme more robustly, the detection mode at P4 needs to be
simplified, and the effects of statistical fluctuation and the
detection noise need to be included in the security analysis. In doing
so, a two-threshold detection illustrated in
Fig.~\ref{Fig:PNA_detector} is used and the position of ``untagged
bits'' is redefined. In Fig.~\ref{Fig:PNA_detector}, through an
integrator, the voltage $v$ denotes the photon number $m$ which is
detected by a common photodiode. Two comparators would output ``11''
when $v\in[v_1,v_2]$,  which means the photon number $m$ at P4 falls
in $[m_1,m_2]$ where $m_1$ and $m_2$ correspond to $v_1$ and $v_2$,
respectively.

To discuss the position and transmittance of ``untagged bits'' in
the passive scheme of a two-threshold detection mode, we consider three
cases according to different parameters of the beam splitter (BS), attenuator, and
detector in Fig.~\ref{Fig:BS_Det}.
\begin{description}
  \item[Case \uppercase\expandafter{\romannumeral 1}~$\left(t_Bt_D=1-t_B\right)$:]
 The ``untagged bits'' are redefined as the photon pulses with photon number $n_5\in[m_1,m_2]$
at P5 in Fig.~\ref{Fig:BS_Det}. Thus, the lower probability bound of
$1-\delta$ $\left(=\sum_{n_5=m_1}^{m_2}P(n_5)\right)$ of the ``untagged
bits'' needs to be estimated. Note that $t_Bt_D=1-t_B$, $P(n_5)$
\big($=B[P(n_1),(1-t_B)]$\big) at P5 is equal to $D(m)$
\big($=B[P(n_1),t_Bt_D]$\big) at P4 in Fig.~\ref{Fig:BS_Det}. Thus,
$1-\delta$ is estimated by $\sum_{m=m_1}^{m_2}D(m)$ and
$\lambda^A=\lambda$.

\begin{figure*}
\centerline{\includegraphics[width=5.5in]{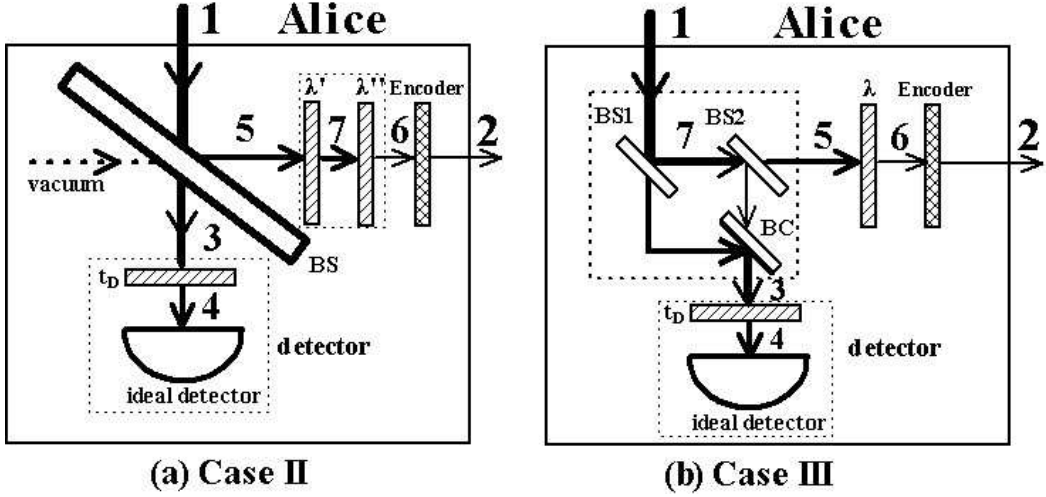}} \caption{The
virtual passive schemes equivalent to Fig.~\ref{Fig:BS_Det} for
Cases II  and III. (a)~Both $\lambda'$
$\left(=t_Bt_D/(1-t_B)\right)$ and $\lambda''$ $(=\lambda/\lambda')$
are the cascaded attenuators which replace $\lambda$ of
Fig.~\ref{Fig:BS_Det}. (b)~BS1: beam splitter (transmittance,
$1-t_Bt_D$); BS2: beam splitter (transmittance, $(1-t_B)/(t_Bt_D)$);
BC: beam combiner. BS1, BS2 and BC replace the BS of
Fig.~\ref{Fig:BS_Det}.} \label{fig_virtual}
\end{figure*}

  \item[Case \uppercase\expandafter{\romannumeral 2}~$\left(\lambda\leq t_Bt_D/(1-t_B)<1\right)$:]
The attenuator $\lambda$ in Fig.~\ref{Fig:BS_Det} is virtually
replaced by two cascaded attenuators $\lambda'$
$\left[=t_Bt_D/(1-t_B)\right]$ and $\lambda''$
$(=\lambda/\lambda')$, in which the security is not reduced. A
virtual passive scheme  [Fig.~\ref{fig_virtual}(a)] is equivalent to
that of Fig.~\ref{Fig:BS_Det} because the PND at P1 is the same as
that at P2, P3, P4, P5 and P6 in both passive schemes. Note that
$\lambda''\leq1$ requires $\lambda\leq t_Bt_D/(1-t_B)$ for the BB84
protocol and $\lambda_d<\lambda_s\leq t_Bt_D/(1-t_B)$ for the decoy-state protocol. In Fig.~\ref{fig_virtual}(a), the ``untagged bits''
are defined as the photon pulses with photon number
$n_7\in[m_1,m_2]$ at P7. Thus, the lower probability bound of
$1-\delta$ $\left(=\sum_{n_7=m_1}^{m_2}P(n_7)\right)$ of ``untagged
bits'' needs to be estimated. Note that $t_Bt_D=(1-t_B)\lambda'$,
$P(n_7)$ at P7 is equal to $D(m)$ at P4 in
Fig.~\ref{fig_virtual}(a). Thus, $1-\delta$ is estimated by
$\sum_{m=m_1}^{m_2}D(m)$ and
$\lambda^A=\lambda''=(1-t_B)\lambda/(t_Bt_D)$.

  \item[Case \uppercase\expandafter{\romannumeral 3}~($t_Bt_D>1-t_B$):]
The BS in Fig.~\ref{Fig:BS_Det} is virtually replaced by two beam
splitters (BS1 and BS2) and a beam combiner (BC)
[Fig.~\ref{fig_virtual}(b)] because the PND at P1  is the same as
that at P2, P3, P4, P5 and P6 in both passive schemes. In
Fig.~\ref{fig_virtual}(b), the ``untagged bits'' are defined as the
photon pulses with photon number $n_7\in[m_1,m_2]$ at P7. Note that
$P(n_7)$ at P7 is equal to $D(m)$ at P4 in
Fig.~\ref{fig_virtual}(b). Thus, the lower probability bound of
$1-\delta$ $\left(=\sum_{n_7=m_1}^{m_2}P(n_7)\right)$ of the ``untagged
bits'' is estimated by $\sum_{m=m_1}^{m_2}D(m)$ and
$\lambda^A=(1-t_B)\lambda/(t_Bt_D)$.

\end{description}

The general form of $\lambda^A$ for the above cases is thus
$(1-t_B)\lambda/(t_Bt_D)$, which is also derived in
\cite{ZQLQ_Untru_09}. Note that, in our method, the lower bound of
$\sum_{m=m_1}^{m_2}D(m)$ should be estimated  by the measured data
from the detection mode shown in Fig.~\ref{Fig:PNA_detector}.

\subsection{PNA without detection noise}
Let the random variable $R=1$ denote that C1 and C2 output ``11'',
otherwise $R=0$. Thus, $R$  follows the binomial distribution
$B\left(1,p\right)$. Without any detection noise in
Fig.~\ref{Fig:PNA_detector}, $p=\sum_{m=m_1}^{m_2}D(m)=1-\delta$.
After $M$ repeating measurements, let the random variable $X=k$
denote $k$ measurements finding $R=1$ and then $X$ follows the
binomial distribution $B\left(M,p\right)$. Statistically, $k/M$
fluctuates around $p$. To estimate the statistical fluctuation, we
give the following Lemma.

\emph{Lemma.} (Clopper-Pearson Confidence
Interval~\rm{\cite{Clopper_Pearson_CI_34,Larsen_CI_86}}) Let $X$ be
the number of successes in $M$ Bernoulli trials with probability $p$
of success on each trial. The Clopper-Pearson ($1-\alpha$)
confidence interval for $p$ is obtained as follows: If $X=x$ is
observed, then the lower and upper bounds $p_l(x,\alpha)$ and
$p_u(x,\alpha)$, respectively, are defined by
\begin{eqnarray}
&&\sum_{j=x}^M{M \choose
j}p_l^j(x,\alpha)\left[1-p_l(x,\alpha)\right]^{M-j}=\frac{\alpha}{2},~
(1\leq x\leq M);\nonumber\\
&&\sum_{j=0}^x{M \choose
j}p_u^j(x,\alpha)\left[1-p_u(x,\alpha)\right]^{M-j}=\frac{\alpha}{2},~(0\leq
x\leq M-1);\nonumber\\
&&p_l(0,\alpha)=0,~p_u(M,\alpha)=1.
\end{eqnarray}

Obviously, from the Lemma, one has
\begin{equation}\label{PNA_Sampling}
{\rm{Prob}}(p_l(k,\alpha)\leq p\leq p_u(k,\alpha))=1-\alpha.
\end{equation}
Thus, $1-\delta$ or $p$ can be lower bounded by $p_l(k,\alpha)$ with
a confidence level $1-\alpha$, while $p_l(k,\alpha)$ can be easily
calculated using the \rm{MATLAB} program.

\subsection{PNA with known and additive detection noise}
In practice, some detection noise exists in
Fig.~\ref{Fig:PNA_detector} and affects the estimation of
$1-\delta$. Here, it is supposed that the properties of detection
noise are known by Alice and are independent of signal detection.
Two kinds of noise are mainly concerned. One is Poissonian noise or
dark counts from the detector itself. The other is Gaussian
electronic noise generated by electronic devices such as the
integrator and comparators in Fig.~\ref{Fig:PNA_detector}. Let the
random variables $m'$, $m$ and $d$ (or $x$) be  measured data, true
photon number, and additive Poissonian noise (Gaussian electronic
noise) and satisfy
\begin{eqnarray}\label{PNA_poisson_additive}
m'=m+d,~m'=m+x.
\end{eqnarray}
For the  Poissonian noise $d$ of the probability
$N(d)=\exp(-\gamma)\gamma^d/d!$, based on
Eq.~(\ref{PNA_poisson_additive}),  one yields
\begin{eqnarray}\label{PNA_Poisson_noise_estimation}
P(m_1\leq m'\leq
m_2)=&&\sum_{m=0}^{m_1-1}D(m)\sum_{d=m_1-m}^{m_2-m}N(d)\nonumber\\
&&+\sum_{m=m_1}^{m_2}D(m)\sum_{d=0}^{m_2-m}N(d).
\end{eqnarray}
Let
\begin{equation}\label{PNA_Poisson_noise_estimation_Para}
\begin{aligned}
\overline{b}(m_1,m_2)&=\max\left\{\sum_{d=m_1-m}^{m_2-m}N(d),m=0,\ldots,m_1-1\right\},\\
b(m_2)&=\sum_{d=0}^{m_2}N(d).
\end{aligned}
\end{equation}
Note that $b(m_2)\geq\overline{b}(m_1,m_2)$. Combining
Eqs.~(\ref{PNA_Poisson_noise_estimation}) and
(\ref{PNA_Poisson_noise_estimation_Para}), one has
\begin{eqnarray}\label{PNA_Poisson_noise_estimation2}
&&P(m_1\leq m'\leq m_2) \leq
\overline{b}(m_1,m_2)\sum_{m=0}^{m_1-1}D(m)
+b(m_2)\sum_{m=m_1}^{m_2}D(m)\nonumber\\
&&\leq \overline{b}(m_1,m_2)\left[\sum_{m=0}^{m_1-1}D(m)+\sum_{m=m_2+1}^\infty D(m)\right]+b(m_2)\sum_{m=m_1}^{m_2}D(m)\nonumber\\
&&=\overline{b}(m_1,m_2)\left[1-\sum_{m=m_1}^{m_2}D(m)\right]+b(m_2)\sum_{m=m_1}^{m_2}D(m),
\end{eqnarray}
which gives
\begin{eqnarray}\label{PNA_Poisson_noise_estimation3}
1-\delta=\sum_{m=m_1}^{m_2}D(m)\geq \frac{P(m_1\leq m'\leq
m_2)-\overline{b}(m_1,m_2)}{b(m_2)-\overline{b}(m_1,m_2)}.
\end{eqnarray}
After $M$ measurements of $m'$, if one finds $k'$ events from the
binomial distribution $B(M,p')$, according to the Lemma,
\begin{equation}\label{PNA_Poisson_Sampling_noise}
{\rm{Prob}}(p_l(k',\alpha)\leq p'\leq p_u(k',\alpha))=1-\alpha,
\end{equation}
where $p'=P(m_1\leq m'\leq m_2)$. From Eqs.~(\ref{PNA_Poisson_noise_estimation3}) and (\ref{PNA_Poisson_Sampling_noise}), $1-\delta$ can be lower
bounded by
\begin{equation}\label{PNA_Poisson_noise_upper}
1-\delta\geq\frac{p_l(k',\alpha)-\overline{b}(m_1,m_2)}{b(m_2)-\overline{b}(m_1,m_2)}
\end{equation}
 with a confidence level $1-\alpha$.

For the case of electronic noise $x$ of the Gaussian probability
$G(x)=(2\pi\sigma^2)^{-1/2}\exp\big[-x^2/(2\sigma^2)\big]$, based on
Eq.~(\ref{PNA_poisson_additive}), one yields
\begin{eqnarray}\label{PNA_Gaussian_noise_estimation}
&&P(m_1\leq m'\leq
m_2)=\sum_{m=0}^{m_1-1}D(m)\int_{m_1-m}^{m_2-m}G(x)dx
\nonumber\\& &+\sum_{m=m_1}^{m_2}D(m)\int_{m_1-m}^{m_2-m}G(x)dx+\sum_{m=m_2+1}^{\infty}D(m)\int_{m_1-m}^{m_2-m}G(x)dx\nonumber\\
&
&\leq\sum_{m=0}^{m_1-1}D(m)\int_{0}^{m_2-m_1}G(x)dx+\sum_{m=m_1}^{m_2}D(m)\int_{-(m_2-m_1)/2}^{(m_2-m_1)/2}G(x)dx
\nonumber\\&&+\sum_{m=m_2+1}^{\infty}D(m)\int_{m_1-m_2-1}^{-1}G(x)dx.
\end{eqnarray}
To derive the above inequality, the property that $G(x)$ is
single-peak function symmetrical at zero is used. Let
\begin{equation}\label{PNA_Gaussian_parameter}
\begin{aligned}
b_1&=\int_{0}^{m_2-m_1}G(x)dx,~
b_2=\int_{-(m_2-m_1)/2}^{(m_2-m_1)/2}G(x)dx,\\
b_3&=\int_{m_1-m_2-1}^{-1}G(x)dx,
\end{aligned}
\end{equation}
and $b_2\geq b_1\geq b_3$. Equation (\ref{PNA_Gaussian_noise_estimation})
then changes to
\begin{eqnarray}\label{PNA_Gaussian_noise_estimation2}
&&P(m_1\leq m'\leq m_2)\nonumber\\&&\leq
b_1\left[\sum_{m=0}^{m_1-1}D(m)\right]+b_2\left[\sum_{m=m_1}^{m_2}D(m)\right]
+b_3\left[\sum_{m=m_2+1}^{\infty}D(m)\right]\nonumber\\
&&\leq
b_1\left[\sum_{m=0}^{m_1-1}D(m)+\sum_{m=m_2+1}^{\infty}D(m)\right]+b_2\left[\sum_{m=m_1}^{m_2}D(m)\right]\nonumber\\
&&=b_1\left[1-\sum_{m=m_1}^{m_2}D(m)\right]+b_2\left[\sum_{m=m_1}^{m_2}D(m)\right].
\end{eqnarray}
Thus, we get
\begin{equation}\label{PNA_Gaussian_noise_estimation3}
1-\delta=\sum_{m=m_1}^{m_2}D(m)\geq\frac{P(m_1\leq m'\leq
m_2)-b_1}{b_2-b_1}.
\end{equation}
Similar to Eq.~(\ref{PNA_Poisson_noise_upper}), $1-\delta$ can be
lower bounded by
\begin{equation}\label{PNA_Gaussian_noise_upper}
1-\delta\geq\frac{p_l(k',\alpha)-b_1}{b_2-b_1}
\end{equation}
 with a confidence level
$1-\alpha$.

\section{Numerical Simulations} \label{simulations}
\begin{figure}
\centerline{\includegraphics[width=4in]{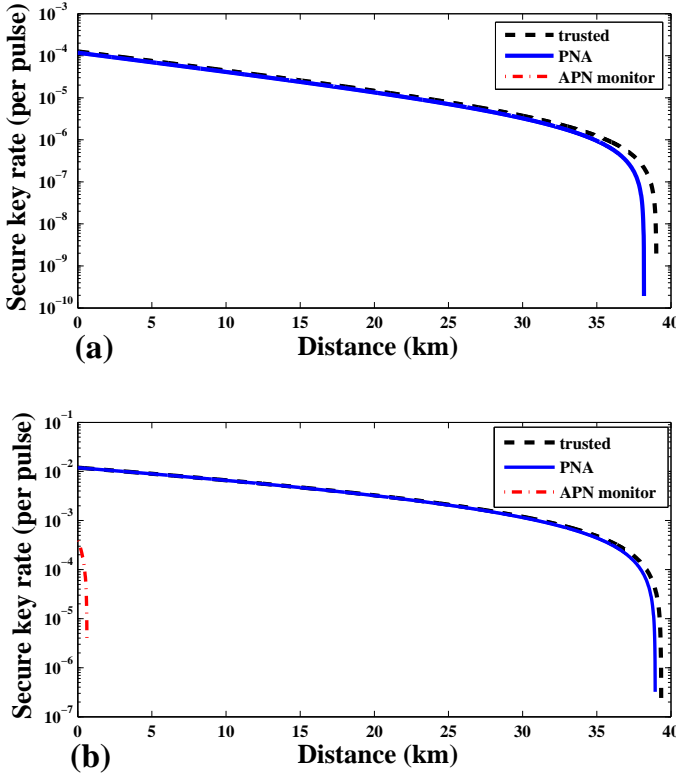}} \caption{(Color
online) The simulations of the BB84 protocol for the trusted source and
untrusted source with the APN monitor and PNA.
(a)~$\eta=(\eta_B\eta_f)/\mu$ and parameters in \cite{GYS_04} are
used. (b)~$\eta=10^{-7}$, $\eta_B=0.5$ and other parameters are the
same as \cite{GYS_04}.} \label{fig_BB84_simulations}
\end{figure}

Based on the BB84 protocol, Fig.~\ref{fig_BB84_simulations} shows the
numerical simulation results for a trusted source and an untrusted source
with an APN monitor and PNA. In the simulations, both the trusted and
untrusted sources are of Poissonian PND. The APN ($\mu$)  of
the untrusted source is $10^6$. We use the passive scheme of $t_B=0.9$
and $t_D=0.76$ to implement the APN monitor and PNA without considering
statistical fluctuation and detection noise.
Figure~\ref{fig_BB84_simulations}(a) shows the result with the
experimental parameters in \cite{GYS_04} (see Table
\ref{tab:para1}), where $\eta_B$ is the efficiency of Bob's
detection, $Y_0$ is the dark-count rate of Bob's detector, and $e_{\rm{det}}$
($e_0$) is the probability that a photon (dark count) hit the
erroneous detector in Bob's side. In
Fig.~\ref{fig_BB84_simulations}(a), we choose the optimized value of
$\eta=(\eta_B\eta_f)/\mu$ for both the trusted and untrusted sources,
$f(E_e)=1$ for perfect error correction and
$[m_1,m_2]=[677160,690840]$  for the detection thresholds in
Fig.~\ref{Fig:PNA_detector}. $Q_e$ and $E_e$ are given by
\cite{Practical_05}
\begin{equation}\label{Simulation_QeEe}
\begin{aligned}
Q_e&=Y_0+1-\exp(-\mu\eta\eta_B\eta_f),\\
E_e&=\frac{e_0Y_0+e_{\rm{det}}[1-\exp(-\mu\eta\eta_B\eta_f)]}{Q_e}.
\end{aligned}
\end{equation}
Note that,  from Fig.~\ref{fig_BB84_simulations}(a), no secure key
rate with an APN monitor is generated at any distance, while the secure key
rate with PNA is found close to the value of the trusted source.
\begin{table}[hbt]
\caption{The simulation parameters for
Fig.~\ref{fig_BB84_simulations}(a).}
\begin{ruledtabular}
\begin{tabular}{cccccccc}
$\mu$&$t_B$&$t_D$ &$\eta_B$&$\alpha'$&$Y_0$&$e_{\rm{det}}$&$e_0$\\
\hline $10^6$&0.9&0.76& 0.045&0.21&$1.7\times10^{-6}$&3.3\%&0.5
\end{tabular}
\end{ruledtabular}
\label{tab:para1}
\end{table}
For comparing the APN monitor with PNA more visibly and fairly,
Fig.~\ref{fig_BB84_simulations}(b) shows the result with fixing
$\eta=10^{-7}$ and resetting $\eta_B=0.5$. Other simulation
parameters are the same as those in Table \ref{tab:para1}. Using
these parameters, one can fix $\overline{P}_{n_2>1}=0.029843$ to
calculate the secure key rate with the APN monitor. From
Fig.~\ref{fig_BB84_simulations}(b), the secure distance is shown to
be less than 1~km with  the APN monitor. Fortunately, PNA improves the
performance which approaches that of the trusted source.

For testing the effects of statistical fluctuation and detection
noise on PNA, we choose an untrusted source of Poissonian statistics
to perform the simulations based on the three-intensity decoy-state
protocol. In the passive scheme as in Fig.~\ref{Fig:BS_Det}, $t_B$ and
$t_D$  are chosen as 0.9 and 0.76, respectively. The APN of
the untrusted source is $1.462\times10^7$. Thus, $\langle m\rangle$ at
P4 is $10^7$. For applying the decoy-state protocol, the average
photon number for a signal (weak decoy) state is $\nu_s=0.5$
($\nu_d=0.1$). In doing so, the transmittance of the attenuator is
$\lambda_s=3.42\times10^{-7}$ ($\lambda_d=6.84\times10^{-8}$) for
the signal (weak decoy) state. The photoelectron detection and
additive detection noise generated in Fig.~\ref{Fig:PNA_detector}
are simulated using the Monte Carlo method and $M=10^8$ of measurements
are run. Other parameters are chosen from  \cite{GYS_04} and
summarized in Table \ref{tab:para2}.
\begin{table*}[hbt]
\caption{The simulation parameters for
Fig.~\ref{fig_PNA_simulations}.}
\begin{ruledtabular}
\begin{tabular}{ccccccccccccc}
$\nu_s$ & $\nu_d$&$t_B$ &$t_D$&$\langle m\rangle$&$M$ & $\lambda_s$ & $\lambda_d$&$\eta_B$&$\alpha'$&$Y_0$&$e_{\rm{det}}$&$e_0$\\
\hline 0.5& 0.1&0.9&0.76&$10^7$&$10^8$ &$3.42\times10^{-7}$ & $6.84\times10^{-8}$&0.045&0.21&$1.7\times10^{-6}$&3.3\%&0.5
\end{tabular}
\end{ruledtabular}
\label{tab:para2}
\end{table*}

\begin{figure}
\centerline{\includegraphics[width=4in]{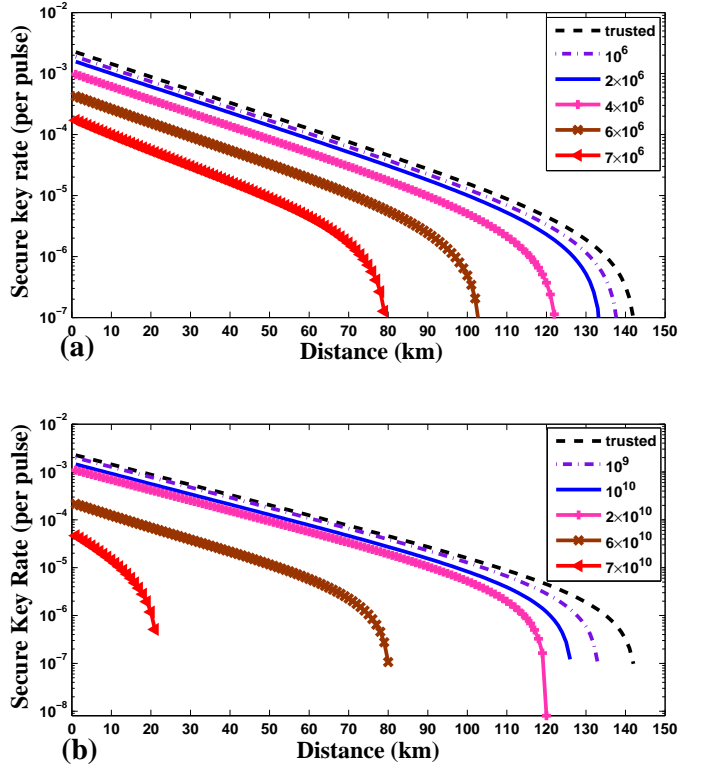}} \caption{(Color
online) The simulations of three-intensity decoy-state protocol for
a trusted source and untrusted source with PNA with different
detection noise, run by $M=10^8$ and $\alpha=10^{-6}$.
(a)~Poissonian noise with the average photon number
$\gamma=10^6,\ldots,7\times10^6$. (b)~Gaussian electronic noise with
variance $\sigma^2=10^9,\ldots,7\times10^{10}$.}
\label{fig_PNA_simulations}
\end{figure}

With different Poissonian noise ($\gamma=10^6,\ldots,7\times10^6$)
added in the PNA, each color line  in Fig.~\ref{fig_PNA_simulations}(a)
shows the secure key rate for the untrusted source.  For comparison, the
black dashed line shows the secure key rate for the trusted source with
the same setup. Here, the confidence level $1-\alpha$ is chosen as
$1-10^{-6}$ and the minimal (maximal) value of detected $m'$ is
chosen as $m_1$ ($m_2$), by which $p_l(M,\alpha)=(\alpha/2)^{1/M}$
can be substituted to Eq.~(\ref{PNA_Poisson_noise_upper}) according
to the Lemma.

With different Gaussian electronic noise
($\sigma^2=10^9,\ldots,7\times10^{10}$) added in PNA, each color
line in Fig.~\ref{fig_PNA_simulations}(b) shows the secure key rate
for untrusted source. The method of choosing $\alpha,~m_1$, and $m_2$
is the same as that of Poissonian noise.

\section{Conclusion} \label{conclusion}
In the passive scheme as in Fig.~\ref{Fig:BS_Det}, a different monitor
mode at P4 gives different statistical characteristics of the untrusted
source and affects the performance of the QKD system. It is shown from
Fig.~\ref{fig_BB84_simulations}  that the PNA can enhance the QKD
performance better than  the APN monitor because PNA utilizes the
two-threshold detection as in Fig.~\ref{Fig:PNA_detector}.
Asymptotically, if the photon statistics at P1 can be fully
characterized through a photon-number-resolving (PNR) detector at P3
and the PND monitor can be realized, the GLLP's analysis for the trusted
source can be efficiently applied and the results are shown as the
black dashed line in
Figs.~\ref{fig_BB84_simulations} and \ref{fig_PNA_simulations}. Without
considering the statistical fluctuation and detection noise, the
secure key rate through PNA approaches that of an ideal PND monitor
except at a long distance. What is more important is that the
two-threshold detection mode of PNA is easier to realize than the PNR
detector. Besides, it is also found from
Fig.~\ref{fig_PNA_simulations} that too large a random noise added in
the PNA's detection would degenerate the QKD performance. Thus, for
improving the secure key rate, the signal-to-noise ratio
(SNR) in Fig.~\ref{Fig:PNA_detector} needs to be kept in an
applicable level. For example, in our simulations,
$R^{SN}_p=\langle m\rangle/\gamma$ and $R^{SN}_g=\langle
m\rangle/\sigma^2$ can be calculated for Poissonian and Gaussian
electronic noise, respectively. From Fig.~\ref{fig_PNA_simulations},
if the QKD distance would exceed 120 km, then $R^{SN}_p\geq2.5$
and $R^{SN}_g\geq5\times10^{-4}$ are needed and is achievable in
practice. Therefore, the passive scheme with PNA is highly practical
to solve the untrusted source problem in the ``Plug \& Play'' QKD
system.

We remark that the effect of parameter fluctuations has not yet been
included in the security analysis. The effective method to deal with
the parameter fluctuations
\cite{Wang_07_source,Wang_Fluc_PRA_07,Wang_08_source,Wang_09_source}
is encouraged to  be applied in the passive scheme.

\emph{Note added.} For passive scheme, the security analysis with
considering statistical fluctuation and detection noise for PNA is
given with using different techniques \cite{ZQLQ_Untru_09}.

\section*{ACKNOWLEDGMENTS}

X. Peng thanks X. X. Lin and Z. Wang for fruitful discussions on the LP
problem. This work is supported by the Key Project of the National
Natural Science Foundation of China (Grant No. 60837004) and the
National Hi-Tech Program of China (863 Program).

\bibliography{Bibli}

\end{document}